\documentstyle[12pt]{article}
\input epsf
\begin{document}
\vskip 0.5cm
\centerline{\large 
Studies of the Vector-Meson Mass Generation Scheme by Chiral}
\centerline{\large 
Anomalies in two-Dimensional non-Abelian Gauge Theories}
\vskip 0.2cm 
%
% authors
\begin{center}
Guanghua Xu$^{\ast}$\\
{\sl University of California, Riverside, California 92521}\\
\end{center}
%
%  Abstract
\begin{abstract}
Higher order effects of 
the two-dimensional non-Abelian gauge theories, in which the 
vector-meson mass is generated by chiral anomalies, will be studied.
The $\beta$ function and the topological nature of the non-linear $\sigma$ 
model in the action and unitarity of the theory will be discussed. 
\end{abstract}
\noindent{$Introduction -$}
The fact that chiral symmetry in a Lagrangian at tree level does not 
survive quantization is well-established, and widely-studied[1], althought we 
still cannot answer the question whether these anomalies have any physical 
significance. There are various ways to remove chiral anomaly[2]. Some leads to 
quite successful prediction[2]. In Jackiw and Rajaraman's work[3], they 
adopted an alternative approach. They considered the two-dimensional chiral 
Schwinger model, and, by simply giving up gauge invariance, obtained a 
consistent and unitary theory, in which the vector meson necessarily acquires 
a mass when consistency and unitarity are demanded. 

R. Rajaraman extended the work to non-Abelian case[6]. At tree level, the 
non-Abelian chiral Schwinger model is invariant under gauge transformations 
$G$, where we can take $G$ as an $SU(N)$ matix without loss of generality.
However, the chiral couplings between the gauge fields and fermionic 
fields will introduce an anomaly. By using the non-Abelian bosonization 
scheme[4], the anomaly will appear in the bosonized action at 
its tree level, and the bosonized form of the model is
\begin{eqnarray}
%eq1
  S_{B}=\sum^{n}_{i=1} S[U_{i}] + \int d^{2}x tr[\frac{1}{2 g^{2}}F^{2}+             
  \frac{i}{2\pi} A_{-} \sum^{n}_{i=1} U_{i}^{\dagger} \partial_{+} U_{i}+
  \frac{n\alpha}{4\pi} A_{+} A_{-}] ,
\end{eqnarray}
where the gauge fields $A^{\pm}=A_{\mp}=({1}/{\sqrt{2}})(A^{0}\pm A^{1})$, 
and $\partial_{\pm}=({1}/{\sqrt{2}})(\partial_{0} \pm \partial_{1})$, and
$S[U_{i}]$ is a Wess-Zumino-Witten action describing a conformally 
invariant nonlinear $\sigma$ model which is equivalent to one flavor of $N$ 
free massless Dirac fermions in two dimensions
%eq2
\begin{eqnarray}
  S[U] = \frac{1}{4\pi} \int d^{2}x tr\partial_{+} U \partial_{-} U^{\dagger}+
  \frac{1}{12\pi} \int_{B} d^{3}y \epsilon^{ijk} trU^{\dagger} \partial_iU 
  U^{\dagger} \partial_j U U^{\dagger} \partial_k U .
\end{eqnarray}
The second term in (2) is the Wess-Zumino term, and is defined
as an integral over a three-dimensional manifold $B$ whose boundary is 
two-dimensional compactified space-time. 

For the gauge field, the gauge transformation for $A_{\pm}^{a}$ is 
$A_{\pm}^{G} = GA_{\pm}G^{+}+i(\partial_{\pm}G)G^{+}$, and for $U_{i}$,
from the bosonization scheme, it is given by $U_{i} \longrightarrow U_{i}G^{+}$.

The last term in (1), a mass term for the gauge field with an arbitrary 
coefficient $\alpha$, reflects an ambiguity in computing the fermionic 
determinant for a given gauge field configuration. The anomaly can not be 
identically zero for any choice of $\alpha$[6].

Following Faddeev's suggestion[6], one can introduce an extra field $V$
in eq. 1 to have 

\vspace{3mm}

 \ \ \ \ \ \ \ \ $S_{B} = \sum^{n}_{i=1} S[U_{i}] + nS[V^{\dagger}] +
  \int d^{2}x tr\{\frac{1}{2g^{2}} F^{2} + \frac{i}{2\pi} A_{-}
  {\sum^{n}_{i=1}}U_{i}^{\dagger} \partial_{+} U_{i}+ $ 

  $\ \ \ \ \ \ \ \ \ \ \ \ \ \ \ \ \ \ 
  \frac{n}{4\pi}[\alpha A_{+} A_{-}+ (\alpha-2)\partial_{+} V^{\dagger}
  \partial_{-} V+i\alpha A_{+} V^{\dagger}\partial_{-} V+i(\alpha-2)A_{-} 
  V^{\dagger}\partial_{+} V]\} .$

\vspace{3mm}

This action is, by design, gauge invariant under 
$A_{\pm} \longrightarrow 
A_{\pm}^{G}$, $U_{i} \longrightarrow U_{i}G^{\dagger}$, and $V^{\dagger} 
\longrightarrow G V^{\dagger}$. In other words, the newly added current 
couples to the gauge fields with opposite chirality from the original current. 
Therefore the anomalies from the newly added scalar fields $V$ cancel those 
of the scalar fields $U$. The gauge choice $V = 1$ reduces it to the original 
system (1), i.e. the new equation is equivalent to eq.(1) and no new degrees
of freedom are introduced as compared to the system in eq.(1).

From ref.[6], we also know that (1) determines a sensible theory, even though
it is not gauge invariant, provided $\alpha$ exceeds 1[6,7].
Contrary to a gauge theory in 1+1 dimensions, which is a trivial one, because
of the anomaly, it has been shown[6,7,8]
that with the regularization $\alpha>1$, the space components $A_{1}^{a}$ of
the vector-field multiplet (and their canonical momenta $E^{a}$) survive as
dynamical variables, in addition to the matter fields.

However the consistence we talked about is only at tree-level (or at
one-loop level for the fermionic theory since the bosonized action (1)
contains fermionic radiative effects and the coefficient $\alpha$
represents regularization ambiguities). Eq.(1) is not completely
solvable. We don't know whether it is still a consistent theory, e.g., 
unitary and hermitian, etc., if perturbation theory is employed. 
Furthermore we don't know whether the theory is renormalizable. In this 
paper, I will look into these matters.    

\noindent{$Quantization - $}In the following, I consider the gauge group
${\cal G} = SU(2)$. But the method is general.
    
By considering the constraints in (1) as second class and replacing all
the Poission brackets by Dirac brackets[10], R. Rajaraman[6,8]
has discussed the quantization of the theory. However, from eq.(1), we
know that the canonical dimension for the vector field operator is
$d(A)=1$. Then by power counting, the renormalizability of (1) is not
straight forward. For the convenience of discussion, I will adopt
Faddeev's proposal[11] to quantize the theory. Introducing an extra
field $V$, where $V = \sqrt{2}(\Psi^{0} \lambda^{0} + i\Psi^{a} \lambda^{a})$ 
is also an element in $SU(2)$ group with $\Psi^{0} \Psi^{0} + 
\Psi^{a} \Psi^{a} = 1, \lambda^{0}=1/{\sqrt{2}}, Tr(\lambda^{a} \lambda^{b})
=\delta^{ab}, [\lambda^{a}, \lambda^{b}] = i f^{abc} \lambda^{c} = i {\sqrt{2
}} {
\epsilon^{abc}} \lambda^{c}$, the enlarged gauge invariant action of eq.(5) 
will be[7,12]
%eq3
\begin{eqnarray}
  S_{B} & = & \sum^{n}_{i=1} \{ (1/{4\pi}) \int d^{2}x\,
  g_{ab}[\vec{\Phi_{i}}] \partial_{\mu} \Phi^{a}_{i} \partial^{\mu}  
  \Phi^{b}_{i}-
  (1/{6\pi}) \int d^{2}x\, \epsilon^{\mu\nu} e_{ab} [\vec{\Phi}_{i}]
  \partial_{\mu} \Phi^{a}_{i} \partial_{\nu} \Phi^{b}_{i}\}+ \nonumber \\
  &   & n \{ [({\alpha-1})/{4\pi}] \int d^{2}x\,
  g_{ab}[\vec{\Psi}] \partial_{\mu} \Psi^{a} \partial^{\mu} \Psi^{b}+
  ({1}/{6\pi}) \int d^{2}x\, \epsilon^{\mu\nu} e_{ab} [\vec{\Psi}]
  \partial_{\mu} \Psi^{a} \partial_{\nu} \Psi^{b} \} + \nonumber \\ 
  &   & \int d^{2}x\, \{- ({1}/{4}) F_{\mu\nu}^{a} F^{a\mu\nu}+
  (n\alpha g^{2}/{8\pi}) A_{\mu}^{a} A^{a\mu}-  \nonumber \\ 
  &   & ({\sqrt{2}} g/{4\pi}) (g^{\mu\nu}+\epsilon^{\mu\nu}) \sum^{n}_{i=1}
  A^{a}_{\mu}[\Phi_{i}^{0} \partial_{\nu} {\Phi_{i}^{a}}- 
  \Phi^{a}_{i} \partial_{\nu} \Phi^{0}_{i}+
  \epsilon^{abc} \Phi_{i}^{b} \partial_{\nu} \Phi_{i}^{c}]- \nonumber \\ 
  &   & ({\sqrt{2}} g n/{4\pi}) [(\alpha-1) g^{\mu\nu}-\epsilon^{\mu\nu}]
  A^{a}_{\mu}[\Psi^{0} \partial_{\nu} {\Psi^{a}}-
  \Psi^{a} \partial_{\nu} \Psi^{0}+
  \epsilon^{abc} \Psi^{b} \partial_{\nu} \Psi^{c}]\} ,
\end{eqnarray}
with[9]
%eq4
\begin{eqnarray}
  g_{ab}[\vec{\Phi}] = \delta_{ab} +(1-{\vec{\Phi}}^{2})^{-1} \Phi_{a} \Phi_{b} ,
\end{eqnarray}
%eq5
\begin{eqnarray}
  e_{ab}[\vec{\Phi}] = \epsilon^{abc} \frac{\pm3 \Phi^c}{2{|\vec{\Phi}|}^{3}} 
  [\arcsin|\vec{\Phi}| - {({\vec{\Phi}}^{2}-{\vec{\Phi}}^{4})}^{1/2}] ,
\end{eqnarray}
where we have let $A^{a}_{\mu} {\,} {\longrightarrow} {\,} gA^{a}_{\mu}$, 
$U = \sqrt{2}(\Phi^{0} \lambda^{0} + i\Phi^{a} \lambda^{a})$.
The action is clearly local gauge invariant under $A_{\pm} \longrightarrow 
A_{\pm}^{G}$, $U_{i} \longrightarrow U_{i}G^{\dagger}$, and $V^{\dagger} 
\longrightarrow GV^{\dagger}$. We see that, upon using the ghost-free gauge 
$V^{\dagger}=1$, one restores 
the original system in eq.(1), and eq.(1) clearly contains the same degrees 
of freedom for the enlarged system (3). Therefore, the action (1) is just 
the action (3) in the gauge $V^{\dagger}=1$ at tree level.

Considering the symmetry of the theory and the simplification of the
discussion, as well as the transparence of renormalizability, I will
choose the covariant, or Lorentz, gauge to fix the gauge invariance
%eq6
\begin{eqnarray}    
  \partial^{\mu} A^{a}_{\mu} =0 .
\end{eqnarray}  
This gauge condition is not ghost-free. The generating functional for the 
theory will be 
%eq7
\begin{eqnarray}
  Z_{B}[{\vec{J}},{\vec{K}},{\vec{L}},{\vec{\alpha}},{\vec{\beta}}] & = & \int 
  [d {\vec{A_{\mu}}}]
  (\prod^{n}_{i=1} [\frac{d {\vec{\Phi_{i}}}}{\sqrt{1-{\vec{\Phi}}^{2}}}])
  [\frac{d {\vec{\Psi}}}{\sqrt{1-{\vec{\Psi}}^{2}}}][d {\vec{c}}]
  [d {\vec{c^{+}}}] exp{\,} i \int d^{2} x \{ {\cal L }_{eff}+ \nonumber \\ 
  &   & J^{a}_{\mu} A^{a\mu} + \sum^{n}_{i=1} K^{a}_{i} \Phi^{a}_{i} +  
  L^{a} \Psi^{a} + c^{a+} \alpha^{a} + \beta^{a} c^{a}\} ,
\end{eqnarray}
where
%eq8
\begin{eqnarray}
  S_{eff} = \int d^{2} x {\cal L }_{eff} = S_{B} - \int d^{2} x \{ 
  (1/{2 \xi}) (\partial^{\mu} A^{a}_{\mu})^{2}+
  c_{a}^{\dagger} \partial^{\mu} [\delta_{ab} \partial_{\mu} -
  g f_{abc} A^{c}_{\mu}] c_{b}(x) \} .
\end{eqnarray}
In eq.(8), $S_{B}$ is given in eq.(3),
$-({1}/{2\xi}) (\partial^{\mu} A^{a}_{\mu})^{2}$ is the gauge-fixing term,
and $c^{a}$ is the Faddeev-Popov(FP) ghost coming from the gauge fixing
condition, and we have introduced the sources $J^{a}_{\mu}$, $K^{a}$, $L^{a}$,
$\alpha^{a}$, and $\beta^{a}$ for the fields $A^{a}_{\mu}$, $\Phi^{a}$, 
$\Psi^{a}$, $c^{a+}$, and $c^{a}$ respectively.

As we can see from the quadratic terms of the effective Lagrangian, 
there are mixings among $\Phi^{a}$, $\Psi^{a}$, and $A^{a}_{\mu}$. From
the quadratic term in eqs.(3,7), we can obtain the propagators for
$(\Phi^{a}, \Psi^{a}, A^{a}_{\mu})$ as $3 \times 3$ matrix $G^{ac}_{ij}$
and the FP ghost propagator. Getting the form of the vertices 
from the interaction terms of the effective 
Lagrangian (8)is straight forward.

Detailed calculations can clearly show that the canonical
dimension for the scalar fields ${\Phi}^{a}, {\Psi}^{a}$ and the gauge 
fields $A^{a}_{\mu}$ are zero in the covariant gauge,in which the theory has 
good high-energy behaviour, therefore the theory is renomalizable. However, we 
also require the renormalized Lagrangian to be gauge-invariant and to have the 
same degrees of freedom in all orders, such that we can be sure the 
equivalence of the theory in different gauges and the renormalizability, 
unitarity, and Lorentz invariance of the action given in (1). 

\noindent{$Regularization$ $and$ $Renormalization - $}
Since we can prove that all the 
one-loop momentum integrals involving odd number of $\epsilon$-tensors are 
finite or can be considered as connected with external lines (see ref.[13] for 
details), we can safely use dimensional regularization at one-loop level, which 
does not destroy the gauge invariance. For two-loop (or higher-order 
calculations), we can discuss them after we obtain the one-loop (or 
corresponding lower-order) renormalized lagrangian (also see ref.[13] for 
details). The additional interaction
of the form $-({1}/{2}) \delta^{2}(0)\int d^{2} x ln[1-{\vec{\Phi}}(x)^{2}]$
from the interaction measure due to the constraint $(\Phi^{0})^{2}+
({\vec{\Phi}}^{a})^{2}=1$ will not affect our discussion[13]. In order to avoid 
the singularities due to the scalar propagators below two dimensions for the 
massless scalar theory and for simplicity of the Ward$-$Takahashi(WT) 
identities coming later, we choose the infrared cutoff as
%eq9
\begin{eqnarray}
  \int d^{2} x[(\pm)H (1-{\vec{\Phi}}^{2})^{1/2} + (\pm)I (1-{\vec{\Psi
  }}^{2})^{1/2} ] ,
\end{eqnarray}
where $H$ and $I$ are arbitrary sources coupled to $\Phi^{0}$ and
$\Psi^{0}$. Indeed the expansions of (12) in powers of ${{\vec{\Phi}}^{2}(x)}$
and ${{\vec{\Psi}}^{2}(x)}$ generate masses for the scalar particles. 
Eventually, $H$ and $I$ will need to take the limit zero.

I shall now assume that the lagrangian has been regularized and discuss
the WT identities and their implications for the structure of the
counterterm. Without loss of generality, I will assume, in the following, that
there is only one flavour (i.e. $n=1$) in the theory.    

The gauge-fixed effective action (8) is obviously not invariant under
the general gauge transformation $G = 1+ i \Omega^{a} {\lambda}^{a}$ with an 
arbitrary $\Omega^{a}$, but it is invariant under the transformation with
$\Omega^{a}=-g c^{a} \delta \lambda$, where $\delta \lambda$ is a Grassmann 
number.

Introducing the source terms $Q^{a}_{\mu}$, ${\nu}^{a}$, ${\delta}^{0}$, 
${\delta}^{a}$, ${\eta}^{0}$, 
${\eta}^{a}$ for the composite operators $(D^{\mu}\vec{c})^{a}$, 
$({g}/{2}) (\vec{c} \times \vec{c})^{a}$, 
$(g\vec{\lambda} \cdot \vec{c} \, U^{\dagger})^{0}$,
$(g\vec{\lambda} \cdot \vec{c} \, U^{\dagger})^{a}$, 
$(g\vec{\lambda} \cdot \vec{c} \, V^{\dagger})^{0}$,     
$(g\vec{\lambda} \cdot \vec{c} \, V^{\dagger})^{a}$ respectively, the generating 
functional will be of the form 
%eq10
\begin{eqnarray}
  Z_{B}[\vec{J}, \vec{K},\vec{L},\vec{\alpha},\vec{\beta},Q^{a}_{\mu},
  \nu^{a}, \delta^{0} , \delta^{a}, \eta^{0}, \eta^{a}] & = & \int \prod_{x} 
  [d{\vec{A}}_{\mu}][\frac{d{\vec{\Phi}}}{(1-{{\vec{\Phi}}^{2}})^{1/2}}] 
  [\frac{d{\vec{\Psi}}}{(1-{\vec{\Psi}}^{2})^{1/2}}]
  [d{\vec{c}}] [d{\vec{c^{\dagger}}}]  \nonumber \\ 
  &   & exp \, i \int d^{2} x \{ {\cal L }_{eff} + 
  \Sigma \} , 
\end{eqnarray}
where ${S}_{eff}$ is given in eq.(8) and 
%eq11
\begin{eqnarray}
  \Sigma & = & J^{a}_{\mu} A^{a\mu}+ K^{a} {\Phi}^{a}+
  L^{a} \Psi^{a}+c^{a \dagger} \alpha^{a}+\beta^{a} c^{a}+
  \delta^{0} (g{\vec{\lambda}} \cdot{\vec{c}} \, U^{\dagger})^{0}+
  \delta^{a}(g{\vec{\lambda}} \cdot{\vec{c}} \, U^{\dagger})^{a}+ \nonumber \\ 
  &   & \eta^{0}(g{\vec{\lambda}} \cdot{\vec{c}} \, V^{\dagger})^{0}+ 
  \eta^{a}(g{\vec{\lambda}} \cdot{\vec{c}} \, V^{\dagger})^{a} + 
  Q^{a}_{\mu} (D^{\mu} {\vec{c}})^{a}
  +{\nu}^{a} (g/2) (\vec{c} \times \vec{c})^{a} - \nonumber \\   
  &   & (H/{2\pi}) (1-{{\vec{\Phi}}^{2}})^{1/2} - (I/{2\pi})
  (1-{\vec{\Psi}}^{2})^{1/2} .
\end{eqnarray}

From $\delta Z=0$ under the transformation $G = 1+ i (-g c^{a} \delta 
{\lambda}) {\lambda}^{a}$, it is not difficult to show 
%eq12
\begin{eqnarray}
  &   & \delta \lambda \int d^{2} x [      
  -\frac{gH}{2\sqrt{2} \pi i} \frac{\delta}{i \delta \delta^{0}}-
  \frac{gI}{2\sqrt{2} \pi i} \frac{\delta}{i \delta \eta^{0}}-
  J^{a}_{\mu} \frac{\delta}{i \delta Q^{a}_{\mu}}+
  \frac{1}{\sqrt{2}} K^{a} \frac{\delta}{i \delta \delta^{a}}+
  \frac{1}{\sqrt{2}} L^{a} \frac{\delta}{i \delta \eta^{a}}+  \nonumber \\ 
  &   & \frac{i}{\xi} \alpha^{a} \partial_{\mu}
  \frac{\delta}{i \delta J^{a}_{\mu}}- 
  \beta^{a} \frac{\delta}{i \delta \nu^{a}}]\,\,
  exp \,\, i \int d^{2} y [ {\cal L }_{eff} + \Sigma ] = 0 . 
\end{eqnarray}
Eq.(11) is the generalized WT identity which relates different types of 
Green's functions.   

The generating functional of the connected Green's functions 
$W[\vec{J},{\vec{K}},\vec{L},\vec{\alpha},\vec{\beta},Q^{a}_{\mu},
  \nu^{a},$ ${\delta}^{0}, {\delta}^{a}, {\eta}^{0}, {\eta}^{a}]=i \, lnZ$ 
of the fields ${\vec{\Phi}}$, $\vec{\Psi}$, $\vec{A}_{\mu}$, 
satisfies the same equation as (11). Defining
%eq13
\begin{eqnarray}
  -\Gamma [ {\vec{a}}_{\mu},  {\vec{\phi}}, \vec{\psi},\vec{C^{\dagger}},
  \vec{C},Q^{a}_{\mu},\nu^{a},\delta^{0}, \delta^{a}, \eta^{0}, \eta^{a}]
  & = & \int d^{2} x [a^{a}_{\mu} J^{a\mu}(x)+ {\phi^{a}} {K^{a}}+
  \psi^{a} L^{a}+ \nonumber \\ 
  &   & C^{\dagger a} {\alpha}^{a}+{\beta}^{a} {C}^{a}]+ W ,  
\end{eqnarray}
where
%eq14
\begin{eqnarray}
  a^{a}_{\mu} & = & -{\delta W}/{\delta J^{a\mu}},\ \ \ \ \ \ \ \ \ \    
  J^{a\mu} = -{\delta \Gamma}/{\delta a^{a}_{\mu}}, \nonumber \\
  \phi^{a} & = & -{\delta W}/{\delta K^{a}},\ \ \ \ \ \ \ \ \ \  
  K^{a} = -{\delta \Gamma}/{\delta \phi^{a} }, \nonumber \\ 
  \psi^{a} & = & - {\delta W}/{\delta L^{a}},\ \ \ \ \ \ \ \ \ \ \ 
  L^{a} = - {\delta \Gamma}/{\delta \psi^{a}}, \nonumber \\ 
  C^{\dagger a} & = & {\delta W}/{\delta \alpha^{a}}, \ \ \ \ \ \ \ \ \ \ \ \ \ 
  \alpha^{a} = - {\delta \Gamma}/{\delta C^{a \dagger}}, \nonumber \\ 
  C^{a} & = & - {\delta W}/{\delta \beta^{a}}, \ \ \ \ \ \ \ \ \ \ \ 
  \beta^{a} = {\delta \Gamma}/{\delta C^{a}} ,
\end{eqnarray}
and setting 
%eq15
\begin{eqnarray}
  \Gamma =\tilde{\Gamma} -\frac{i}{2\xi} \int d^{2} x (\partial_{\mu} a^{a\mu})
  (\partial_{\nu} a^{a\nu}) ,
\end{eqnarray}
then we have the equations for $\tilde{\Gamma}$ from eq.(11) as
%eq16
\begin{eqnarray}    
  \int d^{2} x[ &   & \frac{H}{2\sqrt{2} \pi} \frac{\delta \tilde{\Gamma}}
  {\delta \delta^{0}}+
  \frac{I}{2\sqrt{2} \pi} \frac{\delta \tilde{\Gamma}}{\delta \eta^{0}}+
  i\frac{\delta{\tilde{\Gamma}}}{\delta a^{a\mu}} \frac{\delta{\tilde{\Gamma}}}
  {\delta Q^{a}_{\mu}}+
  \frac{i}{\sqrt{2}} \frac{\delta{\tilde{\Gamma}}}{\delta \phi^{a}} \frac{
  \delta{\tilde{\Gamma}}}{\delta \delta^{a}}+
  \frac{i}{\sqrt{2}} \frac{\delta{\tilde{\Gamma}}}{\delta \psi^{a}} \frac{
  \delta{\tilde{\Gamma}}}{\delta \eta^{a}}+  \nonumber \\ 
  &   & i\frac{\delta{\tilde{\Gamma}}}{\delta C^{a}} \frac{\delta{\tilde{
  \Gamma}}}{ \delta \nu^{a}}]=0 ,
\end{eqnarray}
%eq17
\begin{eqnarray}
  \frac{\delta{\tilde{\Gamma}}}{\delta C^{a \dagger}}+
  \partial_{\mu} \frac{\delta{\tilde{\Gamma}}}{\delta Q^{a}_{\mu}}=0 .
\end{eqnarray}

Performing a loop expansion of the functional $\tilde{\Gamma}$,
the corresponding Feynman diagrams are obtained by expanding the action and 
the integration measure in powers of ${\vec{\Phi}}^{2}$, ${\vec{\Psi}}^{2}$
to the appropriate order. At lowest order, it is not hard to show that
$\tilde{\Gamma}$ is given by
%eq18
\begin{eqnarray}
  {\tilde{\Gamma}}^{(0)} & = & S_{B} + \int d^{x} \{ i c_{a}^{\dagger}(x) 
  \partial^{\mu} [\delta_{ab} 
  \partial_{\mu} - g f_{abc} A^{c}_{\mu}]c_{b} - (H/{2 \pi}) {(1- 
  {\vec{\phi}}^{2})^{1/2}} - \nonumber \\ 
  &   & (I/{2 \pi}) {(1-{\vec{\psi}}^{2})^{1/2}} + 
  Q^{a}_{\mu}(D^{\mu} \vec{c})^{a}
  +\nu^{a} \frac{g}{2} (\vec{c} \times \vec{c})^{a}+ 
  \delta^{0}(g \vec{\lambda} \cdot \vec{c} \, U^{\dagger})^{0}+ \nonumber \\
  &   & \delta^{a}(g \vec{\lambda} \cdot \vec{c} \, U^{\dagger})^{a}+
  \eta^{0}(g \vec{\lambda} \cdot \vec{c} \, V^{\dagger})^{0}+   
  \eta^{a}(g \vec{\lambda} \cdot \vec{c} \, V^{\dagger})^{a} \} .
\end{eqnarray}

For one-loop order, when the cutoff increases, the divergent part of 
${\Gamma}^{(1)}$ is singled out. By adding to the action a counterterm 
$t \cdot [-{\tilde{\Gamma}}^{(1)div} \, + \, O(t)]$,
then a renormalized action $(S+tS_{1})$ can be constructed and 
satisfies the transformation invariance given above.

Noting that ${\tilde{\Gamma}}^{(1)div}$ is a local function of dimension 
$2$ of the $\vec{\Phi}$, $\vec{\Psi}$, $\vec{A_{\mu}}$ fields, and $H(x)$, 
$I(x)$ are also of dimension $2$, by power counting, we can solve the WT 
identities and write the rescaled action in terms of the rescaled fields as  
%eq19
\begin{eqnarray}
  S^{R}_{B} & = & \int d^{2} x \{ (Z_{1} /{4\pi Z_{2}})  
  g^{R}_{ab}[\vec{\phi}] \partial_{\mu} 
  \phi^{a} \partial^{\mu} \phi^{b} - ({Z_{1}}/{6 \pi Z_{3}}) 
  \epsilon^{\mu\nu} e^{ab}_{R} (\vec{\phi})   
  \partial_{\mu} \phi^{a} \partial_{\nu} \phi^{b} ]+ \nonumber \\   
  &   & [({\alpha}^{R} -1) Z_{4} /{4\pi} {Z_{5}}] g^{R}_{ab} \
  [\vec{\psi}] \partial_{\mu} \psi^{a} \partial^{\mu} \psi^{a} + 
  ({Z_{4}}/{6 \pi Z_{6}}) \epsilon^{\mu\nu} e^{ab}_{R} (\vec{\psi})  
  \partial_{\mu} \psi^{a} \partial_{\nu} \psi^{b} -  \nonumber \\ 
  &   & \frac{1}{4} [\partial_{\mu} a^{Rb}_{\nu}- \partial_{\nu}
  a^{Rb}_{\mu} - g^{R} f^{abc} a^{Rb}_{\mu} a^{Rc}_{\nu}]   
  [\partial^{\mu} a^{R b \nu}- \partial^{\nu} a^{R b \mu} - 
  g^{R} f^{bmn} {a^{R m \mu}} {a^{R n \nu}}]+  \nonumber \\ 
  &   & \frac{{\alpha}^{R} (g^{R})^{2}}{8\pi}   
  {a^{R b}_{\mu}} {a^{R b \mu}} - 
  \frac{\sqrt{2}}{4\pi} (\frac{Z_{1}}{Z_{2}} g^{R} g^{\mu \nu} +
  \frac{Z_{1}}{Z_{3}} g^{R} {\epsilon}^{\mu \nu}) a^{R a}_{\mu} [\pm (\frac{1}{
  Z_{1}}- {\vec{\phi}}^{2})^{1/2} \partial_{\nu} \phi^{a}+ \nonumber \\
  &   & (\pm) \phi^{a} \frac{\phi^{m} \partial_{\nu} \phi^{m}}{
  ((1/{Z_{1}})- {\vec{\phi}}^{2})^{1/2}}+
  {\epsilon}^{abc} \phi^{b} \partial_{\nu} \phi^{c}]-
  \frac{\sqrt{2}}{4\pi} [({\alpha}^{R}-1) \frac{Z_{4}}{Z_{5}} g^{R}
  g^{\mu \nu} -
  \frac{Z_{4}}{Z_{6}} g {{\epsilon}^{\mu\nu}}] \times \nonumber \\
  &   & a^{R a}_{\mu} [\pm
  (\frac{1}{Z_{4}}- {{\vec{\psi}}^{2}})^{1/2} \partial_{\nu} \psi^{a}+
  (\pm) \psi^{a} \frac{\psi^{m} \partial_{\nu}
  \psi^{m}}{(({1}/{Z_{4}})- {\vec{\psi}}^{2})^{1/2}}+
  {\epsilon}^{abc} \psi^{b} \partial_{\nu} \psi^{c} ] \, \} - \nonumber \\
  &   & (H/{2 \pi}) [(1/{Z_{1}}) - {\vec{\phi}}^{2} ]^{1/2} -
  (I/{2 \pi}) [(1/{Z_{4}}) - {\vec{\psi}}^{2} ]^{1/2} ,
\end{eqnarray}
and similarly for $S^{R}_{FPG}$, $S^{R}_{GF}$, $Q^{a}_{\mu}$, $\nu^{a}$,
etc., where
%eq20
\begin{eqnarray}
  &   & Z_{1} = 1 + 2t B_{1} , \nonumber \\
  &   & {Z_{1}}/{Z_{2}} = 1-t[\iota - 2 \kappa -2 B_{1} -
  ({4 \pi}/{\sqrt{2} g}) B_{2}] , \nonumber \\
  &   & {Z_{1}}/{Z_{3}} = 1- t[\iota - 2 \kappa -2 B_{1} -
  ({4 \pi}/{\sqrt{2} g}) B_{5}] , \nonumber \\
  &   & Z_{4} = 1+2t B_{3} , \nonumber \\
  &   & \frac{Z_{4}}{Z_{5}} = 1+ t [ \frac{\iota - 2 \kappa}{\alpha -1} + 
  2 B_{3} - \frac{4 \pi}{\sqrt{2} g } \frac{B_{2}}{\alpha -1}] , \nonumber \\
  &   & {Z_{4}}/{Z_{6}} =1- t[\iota - 2 \kappa -2 B_{3} -
  ({4 \pi}/{\sqrt{2} g}) B_{5}] ,  \nonumber \\ 
  &   & a^{R a}_{\mu} = {Z_{7}}^{1/2} a^{a}_{\mu} = [1 + t (\kappa - \iota) ] 
  a^{a}_{\mu} , \nonumber \\ 
  &   & g^{R} = Z_{8} {Z_{7}}^{-3/2} g = [1- 2t (\kappa - \iota) ] g , 
  \nonumber \\ 
  &   & Z_{9} = 1 - ( t \, \iota /2 ) , \nonumber \\ 
  &   & {\alpha}^{R} = \alpha + t [ {\alpha} (2 \kappa - \iota) +
  ({4 \pi}/{\sqrt{2} g }) (B_{2} + B_{4})] , \nonumber \\ 
  &   & g_{ab}^{R} ({\vec{\phi}}) = {\delta}_{ab} + [(1/{Z_{1}})- 
  {\vec{\phi}}^{2}]^{-1} { \phi^{a} \phi^{b} } , \nonumber \\ 
  &   & e^{ab}_{R} = {\epsilon}^{abc} \frac{\pm3 {\phi}^{c}} 
  {2{|\vec{\phi}|}^{3}} 
  [ \arcsin |\vec{\phi}| - |\vec{\phi}| (1/{Z_{1}} - {\vec{\phi}}^{2})^{1/2} ] ,
\end{eqnarray}
and we have assumed the facts of unbroken global gauge symmetry and ghost 
number conservation. Therefore we achieve the renormalization at one-loop 
order.

After this renormalization, the gauge invariance 
of $S_{B}$ remains, but the transformation law of the fields is performed 
under the rescaled fields and coupling constants
%eq21
\begin{eqnarray}
  &   & U^{R} = \pm (1- Z_{1} {\vec{\Phi}}^{2})^{1/2} \lambda^{0} + 
  i {Z_{1}}^{1/2} {\Phi}^{a} \lambda^{a} \longrightarrow U^{R} \, G^{\dagger} ,
  \nonumber \\
  &   & V^{R} = \pm (1- Z_{3} {\vec{\Psi}}^{2})^{1/2} \lambda^{0} + 
  i {Z_{3}}^{1/2} {\Psi}^{a} \lambda^{a} \longrightarrow V^{R} \, G^{\dagger} ,
  \nonumber \\
  &   & {({A^{R}_{\mu}}/{g^{R}})}^{G} \longrightarrow G({A^{R}_{\mu
  }}/{g^{R}}) G^{+} + i ({\partial}_{\mu} G) G^{+} ,   
\end{eqnarray}
with $G$ in the same group representation as the one given before, and the 
integration measure should be modified by ${d \vec{\Phi}}/{[( 
1/{Z_{1}})- {\vec{\Phi}}^{2}]^{1/2}}$ and ${d \vec{\Psi}}/{[(
1/{Z_{4}})- {\vec{\Psi}}^{2}]^{1/2}}$. However these modifications of the 
interaction will only affect the two-loop order. Thus the renormalized 
lagrangian is gauge invariant, and the gauge group and its representation 
remains the same.

For higher loop situation, we can proceed the process inductively. Assuming
that a renormalized action up to order $(n-1)$, which satisfied the WT 
identities, has been constructed, then we can construct the loop expansion 
up to order $n$ with this action. Taking the large cutoff limit, we will 
realize that ${\tilde{\Gamma}}^{(n)div}$ satisfies the same equation as
${\tilde{\Gamma}}^{(1)div}$. The integration of the equation for 
${\tilde{\Gamma}}^{(n)div}$ has already been discussed, and we see that, at 
order $n$, the effects of the renormalization may be again absorbed into 
rescaling of the fields and the coupling constants.

This completes the induction and shows that for this parametrization of the
system, the theory is renormalizable and the renormalized action is gauge
invariant, and the gauge group, its representation, and the degrees of freedom 
remain the same. The renormalized action has the form given by eq.(19). The 
above discussions also leads to the proof of the renormalizability of the 
action (1) and the equivalence between eq.1 and eq.3 at all orders.

\noindent{$More$ $Discussions - $}
The $B_{i}$'s appearing in eq.(20) can be obtained
through perturbative calculation. Renormalization group behaviour and  
topological nature of the non-linear $\sigma$ model in two dimensions has been 
well studied[14]. In the case at hand, due to scalar-vector couplings, the 
affection of these couplings to the $\beta$ function and to the 
topological nature should be re-examined. 

The contribution from the gauge field can come from the propagators
and the scalar-vector couplings. The Feynman diagrams
that could affect the calculation of the $\beta$ function of the
non-linear $\sigma$ model are two- and four-point diagrams of the scalar 
fields. For the diagrams with explicit scalar-vector coupling, since the 
counter terms for them doesn't exist in the tree level Lagrangian, the 
renormalizability of the model would require the sum of them to be finite. 
For the diagrams without explicit scalar-vector coupling, detailed calculations 
show that the contributions from the gauge fields through the propagators are 
finite. Therefore, at least at one-loop order, the scalar-gauge field 
couplings do not affect the $\beta$ function of the model. The $\beta$ function 
behavior of the scalar field interactions, e.g. $\vec{\Phi}$ or $\vec{\Psi}$, 
will be the same as the pure scalar field theories, depending on the 
coefficients in the WZW action. 

Detailed calculations can show that $Z_{1}/Z_{3} = Z_{4}/Z_{6} = 1$, which 
guarantees the topological nature of the nonlinear $\sigma$ model. 

From eq.18, we know that the renormalized action has the same form as eq.3. 
Therefore gauge invariance and the discussion in [6,7,8] guarantee the 
unitarity of eq.1 in higher order, provided ${\alpha}^{R} > 1$. 

Up to this point, higher order effects and gauge invariance of the theory 
have been studied, and the $\beta$ function and the topoplogical nature of the 
nonlinear $\sigma$ model, as well as the untarity of the theory have been 
re-examined. More properties should be further studied. Hope this work
can be one step closer than using the tree-level action to draw conclusion.

This work is supported by the DOE through Grand No. UCR (DE-FG03-86ER40271).

\vspace{7mm}

\noindent{$\ast$ Correspondence address: P25 MS H846,
Los Alamos National Laboratory, Los Alamos,}

\noindent{\,\ \ NM 87545.}

\end{document}